\documentclass[aps,prl,twocolumn,tightenlines,superscriptaddress,nofootinbib]{revtex4-1}
\usepackage{epsfig,rotating,amsmath}
\usepackage{multirow}
\mathchardef\hy="2D
\newcommand{\nuc}[2]{\hbox{$^{#1}$#2}}

\begin{document}
%\draft
\title{Quadrupole collectivity in neutron-deficient Sn nuclei:
  \nuc{104}{Sn} and the role of proton excitations}   

\author{V.\,M.\ Bader}
	\affiliation{National Superconducting Cyclotron Laboratory,
			Michigan State University, East Lansing,
                        Michigan 48824, USA} 
	\affiliation{Department of Physics and Astronomy,
      Michigan State University, East Lansing, Michigan 48824, USA}
\author{A.\ Gade} 
	\affiliation{National Superconducting Cyclotron Laboratory,
			Michigan State University, East Lansing,
                        Michigan 48824, USA} 
  \affiliation{Department of Physics and Astronomy,
      Michigan State University, East Lansing, Michigan 48824, USA}
\author{D.\ Weisshaar}
  \affiliation{National Superconducting Cyclotron Laboratory,
			Michigan State University, East Lansing,
                        Michigan 48824, USA} 
\author{B.\,A.\ Brown}
  \affiliation{National Superconducting Cyclotron Laboratory,
      Michigan State University, East Lansing, Michigan 48824, USA}
  \affiliation{Department of Physics and Astronomy,
      Michigan State University, East Lansing, Michigan 48824, USA}
\author{T.\ Baugher}
  \affiliation{National Superconducting Cyclotron Laboratory,
      Michigan State University, East Lansing, Michigan 48824, USA}
  \affiliation{Department of Physics and Astronomy,
      Michigan State University, East Lansing, Michigan 48824, USA}
\author{D.\ Bazin}
 \affiliation{National Superconducting Cyclotron Laboratory,
      Michigan State University, East Lansing, Michigan 48824, USA}
\author{J.\,S.\ Berryman}
  \affiliation{National Superconducting Cyclotron Laboratory,
      Michigan State University, East Lansing, Michigan 48824, USA}

\author{A.\ Ekstr\"om}
  \affiliation{National Superconducting Cyclotron Laboratory,
      Michigan State University, East Lansing, Michigan 48824, USA}	
	\affiliation{Department of Physics and Center of Mathematics
          for Applications, University of Oslo, N-0316 Oslo, Norway} 
\author{M.\ Hjorth-Jensen}
   \affiliation{National Superconducting Cyclotron Laboratory,
      Michigan State University, East Lansing, Michigan 48824, USA}
	 \affiliation{Department of Physics and Astronomy,	
      Michigan State University, East Lansing, Michigan 48824, USA}
    \affiliation{Department of Physics and Center of Mathematics for
      Applications, University of Oslo, N-0316 Oslo, Norway}
\author{S.\,R.\ Stroberg}
    \affiliation{National Superconducting Cyclotron Laboratory,
      Michigan State University, East Lansing, Michigan 48824, USA}
    \affiliation{Department of Physics and Astronomy,
      Michigan State University, East Lansing, Michigan 48824, USA}
\author{W.\,B.\ Walters}	
    \affiliation{Department of Chemistry and Biochemistry, 
		  University of Maryland, College Park, Maryland 20742, USA}
\author{K.\ Wimmer}
    \affiliation{National Superconducting Cyclotron Laboratory,
      Michigan State University, East Lansing, Michigan 48824, USA}
    \affiliation{Department of Physics,
			Central Michigan University, Mt. Pleasant,
                        Michigan 48859, USA}	  		
\author{R.\ Winkler}
	 \altaffiliation{Present address: Los Alamos National Laboratory,
		Los Alamos, NM 87545, USA.}	
         \affiliation{National Superconducting Cyclotron Laboratory,
			Michigan State University, East Lansing,
                        Michigan 48824, USA}

\date{\today}

\begin{abstract}
We report on the  experimental study of quadrupole collectivity
in the neutron-deficient nucleus \nuc{104}{Sn} using
intermediate-energy Coulomb excitation.
The $B(E2; 0^+_1 \rightarrow 2^+_1)$ value for the excitation of the first 
$2^+$ state in \nuc{104}{Sn} has been measured to be $0.180(37)~e^2$b$^2$
relative to the well-known $B(E2)$ value of \nuc{102}{Cd}.
This result  disagrees by more than one sigma with a recently published
measurement~\cite{Gua13}. Our result indicates that the most modern many-body calculations
remain unable to describe the enhanced collectivity below mid-shell in
Sn approaching $N=Z=50$. We attribute the enhanced collectivity  to proton
particle-hole configurations beyond the necessarily limited
shell-model spaces and suggest the asymmetry of the $B(E2)$-value
trend around mid-shell to originate from enhanced proton excitations
across $Z=50$ as $N=Z$ is approached.     
\end{abstract}

\maketitle
One of the overarching goals of nuclear physics is the development of
a comprehensive model of the atomic nucleus with predictive power
across the nuclear chart. While the structure of nuclei close to
stability is fairly well understood, significant modifications
compared to stable nuclei have been
observed for short-lived rare isotopes with unbalanced numbers of
protons and neutrons. The driving forces behind these structural changes
are manifold, including spin-isospin parts of the nuclear
interaction~\cite{Otsuka13} 
and various facets of many-body correlations~\cite{Forssen13}. Of
particular importance for the development of nuclear
models is experimental data that consistently tracks the effect of
isospin and changed binding, for example. The chain of Sn isotopes has
been a formidable testing ground for nuclear models as some
spectroscopic data is available from $ N=Z=50 $
\nuc{100}{Sn}~\cite{Hinke12} in the proximity of the proton drip-line
to \nuc{134}{Sn}~\cite{Kor00}, beyond the very
neutron-rich doubly-magic nucleus \nuc{132}{Sn}~\cite{Jones10}. In
even-even nuclei, the electromagnetic $B(E2 \uparrow) = B(E2; 0^+_1
\rightarrow 2^+_1)$ excitation strength is a measure of quadrupole
collectivity, sensitive to the presence of shell gaps, nuclear
deformation, and nucleon-nucleon correlations, for example. In the Sn
isotopes, this transition strength has been reported from
\nuc{104}{Sn} to \nuc{130}{Sn}, spanning a chain of 14 even-even Sn
isotopes. The trend is asymmetric with respect to mid-shell and not even the
largest-scale shell-model calculations have been able to describe the
evolution of transition strength across the isotopic chain without
varying effective charges.
In this work, we report the determination of the
\nuc{104}{Sn} $B(E2)$ value from intermediate-energy Coulomb
excitation with
the collision impact parameter restricted to exclude nuclear
contributions.
Our value exceeds the recently published
result~\cite{Gua13} and -- continuing the trend below mid-shell -- is
found at variance with the largest-scale shell 
model calculations. We arrive at a very different conclusion from
~\cite{Gua13} and explain the enhanced collectivity for
neutron-deficient Sn nuclei by considering proton particle-hole
intruder configurations observed in neighboring nuclei and $\alpha$ correlations
towards $N=Z$. It is 
suggested that the interplay of proton intruder configurations and quadrupole
collectivity is a common phenomenon along proton-magic isotope
chains.           

The measurement was performed at the National Superconducting
Cyclotron Laboratory (NSCL) at Michigan State University. The secondary
projectile beam containing \nuc{104}{Sn} and \nuc{102}{Cd} was
produced by fragmentation of a  
140~MeV/u \nuc{124}{Xe} stable primary beam on a
240~mg/cm$^2$ \nuc{9}{Be} production target and separated using a
150~mg/cm$^2$ Al wedge degrader in the A1900 fragment
separator~\cite{a1900}. Despite the total momentum acceptance of the separator
being restricted to 0.41\%, the \nuc{104}{Sn} purity was only 0.006\%. The high
levels of contamination 
originate from low momentum tails of higher rigidity fragments that extend
exponentially  and overlap with the momentum acceptance of the
fragment separator~\cite{Baz08}. Therefore, NSCL's Radio Frequency
Fragment Separator (RFFS) was used for additional filtering.  
Using the time micro structure of the accelerated beam, it deflects
beam particles according to their time-of-flight with a transverse  
Radio Frequency field, effectively leading to a phase
filtering~\cite{Baz09}. The RFFS was tuned to deflect and block part
of the main contaminants, increasing the \nuc{104}{Sn} purity in the
beam cocktail by two orders of magnitude.  At the experimental end
station downstream of the RFFS, the resulting rare-isotope beam was
composed of ~1.25\% \nuc{104}{Sn} and ~2.9\% \nuc{102}{Cd} at average 
rates of 10 and 26 particles per second, for ~\nuc{104}{Sn} and
~\nuc{102}{Cd} respectively.  

About 12.8~m downstream from the RFFS, the secondary beam interacted
with a gold target of 184~mg/cm$^2$ thickness surrounded by the
high-efficiency CsI(Na) scintillator array
(CAESAR)~\cite{caesar} for in-beam $\gamma$-ray spectroscopy. CAESAR
was read out in coincidence with a Phoswich 
detector for particle identification located 0.96~m downstream of the gold
target. CAESAR consists of  
192 CsI(Na) crystals with about 95\% solid-angle coverage. The granularity
of the array enables event-by-event Doppler reconstruction of $\gamma$
rays emitted in flight by the scattered projectiles. The $\gamma$-ray detection
efficiency of the array was determined with \nuc{88}{Y}, \nuc{22}{Na},
\nuc{137}{Cs}, \nuc{133}{Ba} and \nuc{60}{Co} calibration
standards. The measured 
source spectra and efficiencies agree with {\sc Geant4} simulations
which were used to model the in-beam spectral response of CAESAR~\cite{Sim}
(30\% full-energy peak efficiency at 1~MeV~\cite{caesar}). 

\begin{figure}[h]
%        \epsfxsize 7.8cm
%        \epsfbox{fig1.eps}
\includegraphics[width=\columnwidth]{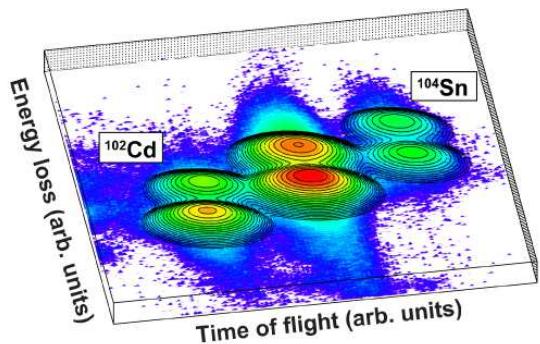}
\caption{\label{fig:pid} (Color online) Particle-identification
  spectrum for the neutron-deficient projectile beam passing through the Au
  target. The energy loss measured in the PIN detector
  is plotted versus the ion's flight time. \nuc{104}{Sn} and
	\nuc{102}{Cd} can be clearly separated.
	The black lines show two-dimensional Gaussian fits.}  
\end{figure}

The incoming projectiles were identified on an
event-by-event basis using energy-loss and time-of-flight
measurements. The ion's energy loss was measured with a 300 $\mu$m Si
PIN detector and the flight time was measured between a plastic
scintillator located at the exit of the A1900 and the Phoswich detector
downstream of CAESAR. The identification is
shown in Fig.~\ref{fig:pid}.

In intermediate-energy Coulomb excitation~\cite{Mot95,Gla98,Gad08},
projectiles are scattered off stable high-$Z$ 
targets and detected in coincidence 
with the de-excitation $\gamma$ rays, tagging the inelastic
process. Peripheral collisions are selected in the regime of
intermediate beam energies to exclude nuclear 
contributions to the otherwise purely electromagnetic excitation
process. This is typically accomplished by restricting the data
analysis to events at very forward scattering angles, corresponding to
large impact parameters, $b>b_{min} =
1.25~$fm$(A_p^{1/3}+A_t^{1/3})+2~$fm, in the interaction between
projectile and target nuclei~\cite{Gad08}. This gives maximum
scattering angles of $\theta_{\rm lab}^{\rm max}=3.126^{\circ}$
and $3.206^{\circ}$ for \nuc{104}{Sn} (67~MeV/u mid-target energy)
and \nuc{102}{Cd} (64~MeV/u mid-target energy), respectively. Assuming
the reaction happens at the beginning of the target, at higher beam
energies, maximum scattering angles of $\theta_{\rm lab}^{\rm
  max}=2.756^{\circ}$ and $2.821^{\circ}$ for \nuc{104}{Sn} and
\nuc{102}{Cd} result. The data
analysis in this work was restricted to $\theta_{\rm lab}^{\rm
  max}=2.578^{\circ}$ for both \nuc{104}{Sn} and \nuc{102}{Cd},
resulting in a conservative minimum impact parameter for both
projectiles. Position measurements from a Parallel Plate Avalanche
Counter (PPAC), placed between CAESAR and the Phoswich detector, were used to 
reconstruct each projectile's outgoing angle. The centering of the
beam on the PPAC  and the focusing on target were carefully
adjusted in a dedicated beam tuning period with two PPACs. With the precisely
known 
distance between PPAC and target, the angles were reconstructed from
the particle's position on the PPAC. A position uncertainty of 0.5~mm
was determined from mask calibrations. This makes the angle measurement based on
geometry more precise than the effect of angular straggling estimated to be
small~\footnote{We estimated the effect of straggling with a Monte Carlo
  simulation that assumes $\sigma=0.4^{\circ}$ angular straggling~\cite{lise}. The input
  distribution was chosen to match the measured angular distribution when folded
  with the straggling distribution. For both, \nuc{104}{Sn} and \nuc{102}{Cd},
  we determined that for all angles larger than $1.5^{\circ}$ about 0.7 to 1.5\%
  of the counts originate from larger scattering angles that fulfill our maximum
  angle cut since they straggled to smaller angles and we lose about 3-4\% of
  events that would have fulfilled the maximum angle cut without angular
  straggling.}.
Systematic uncertainties, including the one related to
straggling, will 
divide out in the normalization procedure outlined below.         

\begin{figure}[h]
%        \epsfxsize 7.8cm
%        \epsfbox{fig2.eps}
\includegraphics[width=\columnwidth]{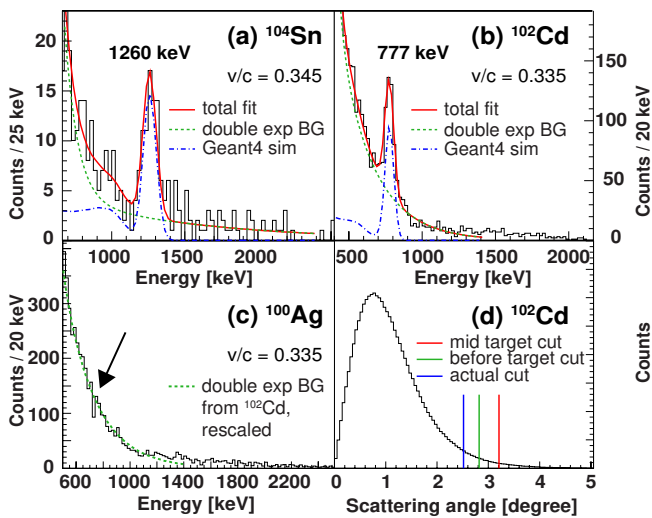}
\caption{\label{fig:gamma} (Color online) Event-by-even Doppler reconstructed
  $\gamma$-ray spectra detected in coincidence with scattered
  \nuc{104}{Sn} (a) and scattered \nuc{102}{Cd} (b).
Gamma-ray transitions at 1260~keV and 777~keV can be clearly
identified and are attributed to the de-excitation $\gamma$ rays from
the first excited $2^+$ state in \nuc{104}{Sn} and in \nuc{102}{Cd},
respectively. (c) shows the \nuc{102}{Cd}
background fit scaled to \nuc{100}{Ag} for which no de-excitations are observed
in the 
sensitive region and (d) displays the scattering angle spectrum with our
conservative angle cut indicated.}   
\end{figure}

The event-by-event Doppler-reconstructed $\gamma$-ray spectra taken in
coincidence  with \nuc{104}{Sn} and \nuc{102}{Cd} respectively  --
with the scattering-angle  restrictions applied -- are shown in
Fig.~\ref{fig:gamma}.  To improve the peak-to-background ratio,
only events with $\gamma$-ray multiplicity 1 were considered.
In agreement with~\cite{ado}, the $\gamma$-ray transition at 1260~keV
observed in \nuc{104}{Sn} and the transition at 777~keV
in\nuc{102}{Cd} are attributed to the decay of the first $2^+$ state
to the $0^+$ ground state. No evidence for other transitions was
observed. 

We would normally determine the absolute angle-integrated
Coulomb excitation 
cross sections, $\sigma(\theta_{\rm lab} \leq \theta_{\rm lab}^{\rm
  max})$, and translate them into absolute $B(\sigma 
\lambda)$ excitation strengths using the Winther-Alder description of
intermediate-energy Coulomb excitation~\cite{Win79}, however, this was not
possible here as the absolute efficiencies of this new 
setup and data acquisition system are not characterized as well as in
our work at the S800 spectrograph (see~\cite{Gad03}, for
example). Therefore, the $B(E2 \uparrow)=B(E2; 0^+ \rightarrow 2^+_1)$
excitation strength for \nuc{104}{Sn} was determined relative to 
the well known $B(E2 \uparrow)$ value of \nuc{102}{Cd}~\cite{Eks09},
which Coulomb excitation yield was measured in the present work at the very same
time, under  
identical conditions, and with good statistics, 
\begin{equation} 
B(E2 \uparrow)^{Sn} = B(E2 \uparrow)^{Cd} \cdot 
\frac{N_{\gamma}^{Sn} N_{proj}^{Cd}
  AW(Cd)}{N_{\gamma}^{Cd}
  N_{proj}^{Sn} AW(Sn)},
\end{equation}  
with $N_{\gamma}^{Sn,Cd}$ and $N_{proj}^{Sn,Cd}$ the number of
$2^+_1$ de-excitation $\gamma$-rays and projectiles for \nuc{104}{Sn}
and \nuc{102}{Cd}, respectively. The term $AW(Sn,Cd)$ is the Alder-Winther
angle-integrated Coulomb excitation cross section, $\sigma(\theta_{\rm lab} \leq
2.578^{\circ})$ per unit $B(E2)$ value, for \nuc{104}{Sn}
and \nuc{102}{Cd}, respectively, taking into account the proper
kinematics and atomic number of the projectiles.  

The number of projectiles, $N_{proj}$, was determined from the
particle identification spectrum shown in
Fig.~\ref{fig:pid}. Two-dimensional Gaussian fits, also shown in the
figure, were used to estimate the contamination from neighboring,
highly intense constituents of the cocktail beam. These corrections to
the number of projectiles were below 3\% for both \nuc{104}{Sn} and
\nuc{102}{Cd}.

The simulated ({\sc Geant4}) response of CAESAR was scaled to the
\nuc{104}{Sn} and \nuc{102}{Cd} data to determine the 
number of $2^+_1$ de-excitations, $N_{\gamma}$. 
The simulations takes into account 
CAESAR's $\gamma$-ray detection efficiency and the 
absorption in the gold target and surrounding materials 
and include the calculated $\gamma$-ray angular
distribution in Coulomb excitation~\cite{Win79,Oll03}. The simulated
response functions fitted on top of a 
double exponential smooth background are shown in
Fig.~\ref{fig:gamma}. 

\begin{figure}[ht]
 %       \epsfxsize 8.2cm
 %       \epsfbox{fig3.eps}
\includegraphics[width=\columnwidth]{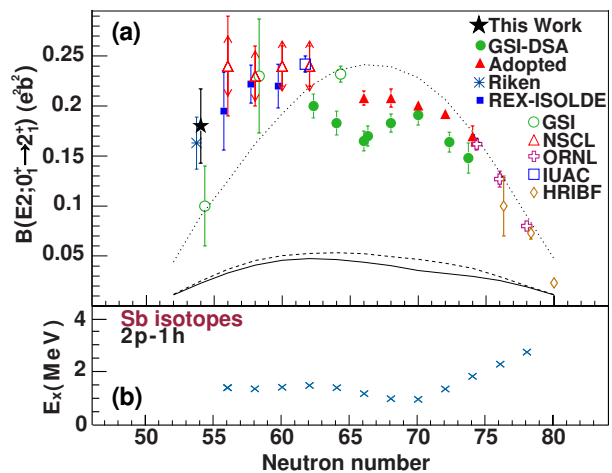}
\caption{\label{fig:evenbe2} (Color online) (a) Measured $B(E2 \uparrow)$
  values for the 
  chain of Sn 
isotopes: Adopted~\cite{ado}, Riken~\cite{Doo13}, REX-ISOLDE~\cite{Ced07,Eks08}, GSI-DSA~\cite{Jun11}, GSI~\cite{Gua13,Banu05,Doo08}, NSCL~\cite{Vam07}, ORNL~\cite{All11}, IUAC~\cite{Kum10}, HRIBF~\cite{Rad05}.    
	Solid line: Large-scale shell-model (LSSM) calculations with
        NNLO interaction \cite{Eks13}.	 
	Dashed line: LSSM with N$^3$LO interaction \cite{entem2003}. The dotted
        line reproduces the gds ($t_{\pi}=4$) calculation shown in~\cite{Banu05}. (b)
        Energies of the proton $2p-1h$ intruder states in Sb
        nuclei~\cite{sb111}.} 
\end{figure}

With the known value of $B(E2 \uparrow)=0.28(3)~e^2$b$^2$ for
\nuc{102}{Cd}~\cite{Eks09}, we deduce $B(E2
\uparrow)=0.180(37)~e^2$b$^2$ for \nuc{104}{Sn}. The uncertainty
includes all statistical uncertainties of $N_{proj}$,
uncertainties from the fits of the response functions used to derive
$N_{\gamma}$ and the uncertainty of the
\nuc{102}{Cd} $B(E2 \uparrow)$ value. 
The aforementioned normalization eliminates systematic
uncertainties stemming from the angle-determination with the PPAC and
absolute efficiencies in general. 
Our result compares well with
the work by Doornenbal~\cite{Doo13} and disagrees with the recently
published value from a relativistic Coulomb
excitation measurement performed at
GSI~\cite{Gua13}. Ref.~\cite{Doo13} discusses 
the impact of unobserved feeding and we point out that, \nuc{102}{Cd} and
\nuc{104}{Sn} have rather comparable proton 
separation energies, $S_p=5.614$~MeV and $S_p=4.286$~MeV, 
respectively,  with a  similar potential of unobserved feeding from
higher-lying $3^-$ and $2^+_n$ states. Nuclear
contributions are minimized in our work by very conservative minimum impact
parameters that the analysis is restricted to, avoiding
model-dependent estimates of the nuclear contribution. Our results and the 
literature values are summarized in Table~\ref{tab:table1}. An
overview of the measured $B(E2 \uparrow)$ values along the Sn isotopic
chain is given in Fig.~\ref{fig:evenbe2}(a).

\begin{table}
\caption{\label{tab:table1} $B(E2; 0^+_1 \rightarrow 2^+_1)$ values for
\nuc{102}{Cd} and \nuc{104}{Sn} from literature and from this work
(\nuc{104}{Sn}); we used the results from~\cite{Eks09} for
normalization in eq.~(1).} 
\begin{ruledtabular}
\begin{tabular}{cccc}
 Isotope&
 \begin{tabular}{@{}c@{}}$B(E2
   \uparrow)_{literature}$\\($e^2$b$^2$)\end{tabular}&  
 Ref & 
 \begin{tabular}{@{}c@{}}$B(E2 \uparrow)$\\($e^2$b$^2$)\end{tabular} \\ 
\hline
\nuc{102}{Cd}&\begin{tabular}{@{}c@{}}0.28(3)\footnotemark[1]\\0.281(45)
\end{tabular} &
							\begin{tabular}{@{}c@{}}\cite{Eks09}\\
\cite{Boe07}\end{tabular}&\\ \hline
\nuc{104}{Sn}&\begin{tabular}{@{}c@{}}0.10(4)\\0.163(26)\end{tabular} &
							\begin{tabular}{@{}c@{}}\cite{Gua13}\\
\cite{Doo13}\end{tabular}&0.180(37)\\

\end{tabular}
\end{ruledtabular}
\footnotetext[1]{The error includes statistical and systematic uncertainties.}
\end{table}

In order to understand the trend in $B(E2 \uparrow)$ strength toward $^{100}$Sn,
we have performed several large-scale shell-model (LSSM) calculations, using a
recently parametrized nucleon-nucleon force based on chiral perturbation theory
\cite{Eks13} (at next-to-next leading order, NNLO) as well as  the
N$^3$LO nucleon-nucleon interaction of~\cite{entem2003}. We have also
studied the influence of three-body forces following Ref.~\cite{hagen2012b}. We
employ $^{100}$Sn as a closed-shell core, 
defined by the quantum numbers of 
the $0h_{11/2}$, $1d$, $2s_{1/2}$ and $0g_{7/2}$ single-neutron
states. One major problem with this model space is that the
single-particle 
energies of $^{101}$Sn are not known experimentally, except for the
spacing between the $7/2_1^{+}$ and 
the $5/2_1^{+}$ states of 170~keV
\cite{darby2010,seweryniak2007}. The effective neutron charge was set
to $0.5e$ for all calculations. 

All of these calculations result in rather similar
behaviors for the $B(E2 \uparrow)$ transitions in the
neutron-deficient tin isotopes. 
In Fig.~\ref{fig:evenbe2} we thus present only the
results for the newly optimized  
NNLO interaction \cite{Eks13}, together with those obtained with the
N$^3$LO interaction \cite{entem2003}. The latter interaction also gives the
overall best reproduction of the excited 
states and binding energies.  Thus, unless one adopts a
phenomenological adjustment of the effective neutron
charges, see for example Ref.~\cite{back2013}, theory based on an
inert proton core fails to describe the $B(E2 \uparrow)$ strengths. 
Our results are similar to the $t=0$ results shown in Fig.~3 of \cite{Banu05}.
If we increase the neutron effective charge to $1.0e$, then the $B(E2)$ values 
are increased by a factor of four and are in better agreement with the 
data. But they are still too small for the neutron-deficient tin isotopes.
   
In \cite{Banu05}, the model space was increased to allow up to four
protons to be excited from the $0g_{9/2}$ orbital to $1d$, $2s_{1/2}$ and
$0g_{7/2}$, ($t_{\pi}=4$). The results from this calculation, with effective
charges $1.5e$ for protons and $0.5e$ for neutrons, are also shown in
Fig.~\ref{fig:evenbe2}. 
Overall, the data is much better described, except that the extended
calculation is symmetric around the middle, whereas experiment
shows an asymmetry with an enhancement at the neutron-deficient end.

In \cite{Vam07}, the comparison to data and theory for the
nickel isotopes was discussed. $^{56}$Ni and $^{100}$Sn are similar with
regard to their structure and shell gaps. At the closed-shell
limit, they are both $jj$ closed shells with the $0f_{7/2}$ and $0g_{9/2}$
orbitals filled, respectively. There are low-lying one-particle
one-hole ($1p-1h$) excitations across the shell gap that give a moderately
large $B(E2)$ strength in $^{56}$Ni (experimentally observed in \cite{Yur04})
and $^{100}$Sn (not yet experimentally observed but calculated for Fig. 4 of
\cite{Gua13}). These $1p-1h$ proton excitations couple 
coherently to the neutron configurations. The resulting increase in the 
$B(E2)$ values could be interpreted in terms of an additional neutron
core-polarization charge for a model space that involves only the valence
neutrons. 

The $^{100}$Sn effective single-particle energy (ESPE) shell gap for the
calculations given was about 7.5 MeV (Fig. 2 in \cite{Banu05}).
In comparison, the ESPE gap for the calculations used for the nickel isotopes in
Fig.~4 of \cite{Vam07} is 7.1~MeV. The correlated gap obtained from the full
$pf$-shell calculation and from experimental ground-state binding energies of
$^{55,56,57}$Ni is 6.4 MeV. The correlated gap for $^{100}$Sn is not known
experimentally and the theoretical value is not provided in \cite{Banu05}.
The shell gaps obtained with some Skyrme energy-density functionals for
$^{100}$Sn are 5.2 MeV (Skx) \cite{skx}, 
5.9 MeV (SLy4) \cite{sly4} and 5.8 MeV (SkM*) \cite{skms}, and for $^{56}N$i are
4.1 MeV (Skx), 4.8 MeV (SLy4) and 4.7 MeV (SkM*). Thus, the shell gaps for
$^{56}$Ni and $^{100}$Sn are similar.

In contrast to the ($t_{\pi}=4$) calculation for tin isotopes that
is symmetric around midshell, the 
equivalent calculation for nickel (Fig.~4 in \cite{Vam07}) shows an asymmetry
with larger $B(E2)$ values at the neutron-deficient end. For nickel, it is
possible 
to carry out a full $pf$ shell calculation and this enhances the
asymmetry. It is likely that the origin of 
the difference between the calculations for nickel and tin 
comes from the truncations that must be made for tin. 

The energy of the two-proton excitation is much lower than
twice the shell gap energy due to pairing and $\alpha$ 
correlations. For $^{102}$Sn, the two excited protons go into the same deformed
quantum states that are occupied by the two valence neutrons.
This lowers the energy due to the $\alpha$-correlation energy 
and increases mixing with the neutron configurations
near $N=Z$. The $\alpha$ correlation persists for
$^{104}$Sn and gradually decreases for increasing
number of neutrons since the configurations for the added
neutrons differ from the proton configurations.
Thus, $\alpha$ clustering provides a mechanism to generate
an asymmetry in the $B(E2)$ values between $N=50$ and $N=82$.
Part of the $\alpha$ clustering could be missed in the ($t_{\pi}=4$)
calculations due to the seniority truncation that had
to be made for tin.  In addition, there is an asymmetry-inducing  
contribution to the $B(E2)$ values beyond ($t_{\pi}=4$) that
is obtained in the nickel calculations.

The energy of the low-lying intruder states with $mp-nh$ configurations is also
sensitive to these correlations. The
$2p-2h$ state in $^{60}$Ni is observed at 3.32~MeV from the
$^{56}$Fe(\nuc{6}{Li},$d$) reaction \cite{li6d}. The calculated energy of this
state in the full $pf$ shell model is in agreement with experiment (see Fig. 20
of \cite{Hon04}). In the $^{100}$Sn region, $2p-1h$ states have been observed
down to $N=56$ in the Sb chain as shown in Fig.~\ref{fig:evenbe2}(b). The
energies for the $2p-1h$ $9/2^+$ levels have been identified 
and described in~\cite{Shroy79} for \nuc{113-119}{Sb} nuclei and extended to
\nuc{107,109}{Sb} in~\cite{Schnare96}.  

For nuclei
closer to $^{100}$Sn nothing is known yet. Observation of these states and a
comparison to the predictions of the models that include proton excitations
are required to complete the understanding of the asymmetry.
This prevalence of proton core excitations near $N=Z$ is also important
for understanding the $B(E2)$ values in  $^{18}$O, $^{42}$Ca and $^{44}$Ca
\cite{gg67_1,gg67_3} -- making the situation encountered in the Sn
isotopes consistent with observations along proton-magic chains
across the nuclear chart.

In summary, we have determined the \nuc{104}{Sn} $B(E2; 0^+ \rightarrow
2^+_1)$ strength from intermediate-energy Coulomb excitation.  
Our result is at variance with a recently published
measurement~\cite{Gua13}. Unlike the conclusion of~\cite{Gua13}, the
departure from large-scale shell-model 
calculations persists for the neutron-deficient Sn isotopes approaching
\nuc{100}{Sn}. We propose this deviation from shell model to
originate from the interplay of proton particle-hole configurations
beyond the necessarily limited shell-model spaces and argue that their
effect on quadrupole collectivity is a common phenomenon
along proton-magic isotope chains.

\begin{acknowledgments}
This work was supported by
the National Science Foundation under Grants No. PHY-1102511,
PHY-0606007, and PHY-1068217 and by the US Department of Energy,
Office of Nuclear Physics, under Grants No. DE-FG02-08ER41556 
and DE-FG02-94ER40834 (U.M.).  Support by the
Research Council of Norway under contract 
ISP-Fysikk/216699 is acknowledged. This research used computational
resources of the Notur project in Norway.
\end{acknowledgments}

\end{document}